\documentclass[12pt]{article}
\usepackage{graphicx}
\usepackage{epsfig}
\begin {document}
\begin{center}
\bf
MIDRAPIDITY PRODUCTION OF SECONDARIES IN pp COLLISIONS AT
RHIC AND LHC ENERGIES IN THE QUARK--GLUON STRING MODEL

G.H. Arakelyan$^*$, C. Merino, C. Pajares, and Yu. M. Shabelski$^{**}$ \\
\vspace{.5cm}
Departamento de F\'\i sica de Part\'\i culas, Facultade de F\'\i sica, \\ 
and Instituto Galego de Altas Enerx\'\i as (IGAE), \\ 
Universidade de Santiago de Compostela, Galicia, Spain \\
E-mail: merino@fpaxp1.usc.es, pajares@fpaxp1.usc.es

\vspace{.2cm}
 
$^{*}$ Permanent address: Yerevan Physics Institute, Armenia \\
E-mail: argev@mail.yerphi.am

\vspace{.2cm}

$^{**}$ Permanent address: Petersburg Nuclear Physics Institute, \\
Gatchina, St.Petersburg 188350 Russia \\
E-mail: shabelsk@thd.pnpi.spb.ru
\vskip 0.5 truecm

A b s t r a c t
\end{center}

We consider the phenomenological implications of the assumption that baryons 
are systems of three quarks connected through gluon string junction. The 
transfer of baryon number in rapidity space due to the string junction 
propagation is considered in detail. At high energies this process leads to a 
significant effect on the net baryon production in $hN$ collisions at
mid-rapidities. The numerical results for midrapidity inclusive densities of 
different secondaries in the framework of the Quark--Gluon String Model are in 
reasonable agreement with the experimental data. One universal value 
$\lambda \simeq 0.25$ for the strangeness suppression parameter correctly 
describes the yield ratios of $\Lambda/p$, $\Xi/\Lambda$, and $\Omega/\Xi$. 
The predictions for $pp$ collisions at LHC energies are also presented.

\vskip 1.5cm


\newpage

\section{Introduction}

The Quark--Gluon String Model (QGSM) and the Dual Parton Model (DPM) are
based on the Dual Topological Unitarization (DTU) and they describe quite
reasonably many features of high energy production processes in both
hadron--nucleon and hadron--nucleus collisions [1--6]. High energy 
interactions are considered as taking place via the exchange of one or several 
Pomerons, all elastic and inelastic processes resulting from cutting through 
or between Pomerons \cite{AGK}. Inclusive
spectra of hadrons are related to the corresponding fragmentation
functions of quarks and diquarks, which are constructed using the
Reggeon counting rules \cite{Kai}.

In the string models, baryons are considered as configurations consisting of 
three  connected strings (related to three valence quarks) called  string 
junction  (SJ) [9--12]. In the processes of secondary production the SJ 
diffusion in rapidity space leads to significant differences in the yields of 
baryons and antibaryons in the midrapidity region even at very high energies 
\cite{Sh1}. 
 
A quantitative theoretical description of the baryon number transfer via SJ 
mechanism was suggested in the 90's. The later experimentally observed 
$p/\bar{p}$ asymmetry at HERA energies was predicted \cite{KP1} and in 
\cite{Bopp} it was noted that the $p/\bar{p}$ asymmetry measured at HERA can 
be obtained by simple extrapolation of ISR data.

Important results on the baryon number transfer due to SJ diffusion in 
rapidity space were obtained in \cite{ACKS} and following papers [17--21]. 

In the present paper, we calculate the inclusive densities of different
secondaries and compare them with recent RHIC data \cite{abe} for $pp$ 
collisions at $\sqrt{s} =$ 200 GeV. The predictions for secondary production 
at LHC energies are also given.

\section{Baryon as \boldmath$3q+SJ$ system}

In QCD, the hadrons are composite bound state configurations built up
from the quark $\psi_i(x), i = 1,...N_c$, and gluon $G^\mu_a(x)$,
$a=1,...,N_c^2-1$, fields. In the string models the colour part of a baryon 
wave function reads as follows (see Fig.~1) \cite{Artru,RV}:

\begin{figure}[htb]
\centering
\includegraphics[width=.5\hsize]{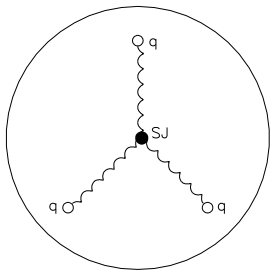}
\vskip -.3cm
\caption{\footnotesize
Composite structure of a baryon in string models. Quarks are shown by open 
points.}
\end{figure}

\begin{eqnarray}
&&B\ =\ \psi_i(x_1) \, \psi_j(x_2) \, \psi_k(x_3) \, J^{ijk}(x_1, x_2, x_3, x) 
\,,
\\
&& J^{ijk}(x_1, x_2, x_3, x) =\ \Phi^i_{i'}(x_1,x) \, \Phi_{j'}^j(x_2,x) \,
\Phi^k_{k'}(x_3,x) \, \epsilon^{i'j'k'} \,,
\\
&& \Phi_i^{i'}(x_1,x) = \left[ T\exp \left(g \cdot \int\limits_{P(x_1,x)}
A_{\mu}(z) dz^{\mu}\right) \right]_i^{i'} \,,
\end{eqnarray}
where $x_1, x_2, x_3$ and $x$ are the coordinates of valence quarks and SJ, 
respectively and $P(x_1,x)$ represents a path from $x_1$ to $x$ which looks 
like an open string with ends at $x_1$ and $x$.

The baryon wave function in Eq.(1) can be defined as a star (or Y) 
configuration. The Y baryon structure is supported by lattice calculations 
\cite{latt}.

This picture leads to some general phenomenological predictions. In
particular, it opens room for exotic states, such as the multiquark bound 
states, 4-quark meson and pentaquark \cite{RV,DPP1,RSh}. In the case of 
inclusive reactions the baryon number transfer to large rapidity distances in 
hadron--nucleon and hadron--nucleus reactions can be explained by SJ diffusion.

\section{Inclusive spectra of secondary hadrons \newline in the
Quark--Gluon String Model}

To perform more quantitative predictions a model for multiparticle production 
has to be adopted. In the present paper we have used the QGSM for the 
numerical calculations. As it was mentioned above, the high energy 
hadron--nucleon collisions are considered in the QGSM as going via the 
exchange of one or several Pomerons. Each Pomeron corresponds to a cylindrical 
diagram (see Fig.~2a), and thus, when cutting a Pomeron, two showers of 
secondaries are produced as it is shown in Fig.~2b. The inclusive spectrum of 
a secondary hadron $h$ is then determined by the convolution of the diquark, 
valence quark, and sea quark distributions $u(x,n)$ in the incident particles 
with the fragmentation functions $G^h(z)$ of quarks and diquarks into the 
secondary hadron $h$. Both the diquark and the quark distribution functions
depend on the number $n$ of cut Pomerons in the considered diagram.

\begin{figure}[htb]
\centering
\includegraphics[width=.6\hsize]{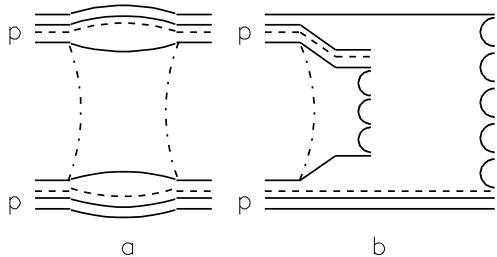}
\vskip -.5cm
\caption{\footnotesize
Cylindrical diagram corresponding to the one--Pomeron exchange contribution to 
elastic $pp$ scattering (a) and the cut of this diagram which determines the 
contribution to the inelastic $pp$ cross section (b). Quarks are shown by 
solid curves and SJ by dashed curves.}
\end{figure}

For a nucleon target, the inclusive spectrum of a
secondary hadron $h$ has the form \cite{KTM}:
\begin{equation}
\frac{dn}{dy}\ = \
\frac{x_E}{\sigma_{inel}} \frac{d\sigma}{dx_F}\ =\ \sum_{n=1}^\infty
w_n\phi_n^h (x)\ ,
\end{equation}
where the functions $\phi_{n}^{h}(x)$ determine the contribution of
diagrams with $n$ cut Pomerons and $w_n$ is the relative weight of
this diagram. Here we neglect the contribution of diffraction
dissociation processes which is very small in the midrapidity region.

For $pp$ collisions
\begin{eqnarray}
\phi_{pp}^h(x) &=& f_{qq}^{h}(x_+,n)f_q^h(x_-,n) +
f_q^h(x_+,n)f_{qq}^h(x_-,n)
\nonumber\\
&&\hspace*{3.5cm}+\ 2(n-1)f_s^h(x_+,n)f_s^h(x_-,n)\ ,
\\
x_{\pm} &=& \frac12\left[\sqrt{4m_T^2/s+x^2}\ \pm x\right] ,
\end{eqnarray}
where $f_{qq}$, $f_q$, and $f_s$ correspond to the contributions of diquarks, 
valence quarks, and sea quarks, respectively.

These functions are determined by the convolution of the diquark and
quark distributions with the fragmentation functions, e.g.
\begin{equation}
f_q^h(x_+,n)\ =\ \int\limits_{x_+}^1u_q(x_1,n)G_q^h(x_+/x_1) dx_1\ .
\end{equation}
The diquark and quark distributions, which are normalized to unity, as well 
as the fragmentation functions are determined by Regge intercepts \cite{Kai}.

At very high energies both $x_+$ and $x_-$ are negligibly small in the 
midrapidity region. In this case all fragmentation functions, which are 
usually written \cite{Kai} as $G^h_q(z) = a_h (1-z)^{\beta}$, are constants, 
\begin{equation}
G_q^h(x_+/x_1) = a_h \ , 
\end{equation}
and lead, in agreement with \cite{AKM}, to
\begin{equation}
\frac{dn}{dy}\ = \ g_h \cdot (s/s_0)^{\alpha_P(0) - 1}
\sim a^2_h \cdot (s/s_0)^{\alpha_P(0) - 1} \,,
\end{equation}
corresponding to the only one-Pomeron exchange diagram in Fig.~3, which is the 
only diagram contributing to the inclusive density in the central region (AGK 
theorem \cite{AGK}). The intercept of the supercritical Pomeron 
$\alpha_P(0) = 1 + \Delta$, $\Delta = 0.139$ \cite{Sh}, is used in the 
numerical calculations. 

\begin{figure}[htb]
\centering
\vspace{-1.5cm}
\includegraphics[width=.55\hsize]{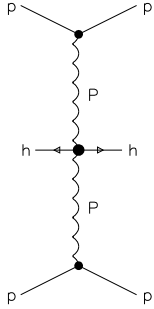}
\vskip -0.1cm
\caption{\footnotesize
One-Pomeron-pole diagram determining secondary hadron $h$ production.}
\end{figure}

The diagram in Fig.~3 predicts equal inclusive yields for each particle and 
its antiparticle. However, some corrections to the spectra of secondary
baryons appear for processes which present SJ diffusion in rapidity space.
Although these corrections would become negligible at energies asymptotically 
high, they result in a significant difference of the baryon and antibaryon 
yields in the midrapidity region for the currently available energy range.
Moreover, this difference vanishes only very slowly when the energy increases. 

According to \cite{ACKS}, we consider three different possibilities to obtain
the net baryon charge. The first one is the fragmentation of the diquark 
giving rise to a leading baryon (Fig.~4a). A second possibility is to produce 
a leading meson in the first break-up of the string and a baryon in the 
subsequent break-up \cite{Kai,22r} (Fig.~4b). In these two cases the baryon 
number transfer is possible only for short distances in rapidity. In the third 
case shown in Fig.~4c both initial valence quarks recombine with sea
antiquarks into mesons $M$ and a secondary baryon is formed by the SJ together 
with three sea quarks.

\begin{figure}[htb]
\centering
\includegraphics[width=.55\hsize]{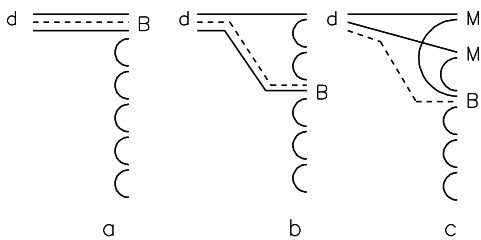}
\vskip -0.1cm
\caption{\footnotesize
QGSM diagrams describing secondary baryon $B$ production by diquark $d$: 
initial SJ together with two valence quarks and one sea quark (a), initial SJ 
together with one valence quark and two sea quarks (b), and initial SJ 
together with three sea quarks (c).}
\end{figure}

The corresponding fragmentation functions for the secondary baryon $B$ 
production can be written as follows (see \cite{ACKS} for more details):
\begin{eqnarray}
G^B_{qq}(z) &=& a_N v_{qq} \cdot z^{2.5} \;,
\\
G^B_{qs}(z) &=& a_N v_{qs} \cdot z^2 (1-z) \;,
\\
G^B_{ss}(z) &=& a_N \varepsilon v_{ss} \cdot z^{1 - \alpha_{SJ}} (1-z)^2
\end{eqnarray}
for the processes shown in Figs.~4a, 4b, and 4c, respectively and where $a_N$ 
is the normalization parameter, and $v_{qq}$, $v_{qs}$, $v_{ss}$ are the 
relative probabilities for different baryons production that can be found by
simple quark combinatorics \cite{AnSh,CS}. The fraction $z$ of the incident 
baryon energy carried by the secondary baryon decreases from Fig.~4a to 
Fig.~4c, whereas the mean rapidity gap between the incident and secondary 
baryon increases. The first two processes can not contribute to the inclusive 
spectra in the central region, but the third contribution is essential if the
value of the intercept of the SJ exchange Regge-trajectory, $\alpha_{SJ}$, is 
close to unity. The contribution of the graph in Fig.~4c has a coefficient 
$\varepsilon$ which determines the small probability of such baryon number 
transfer. 

In \cite{ACKS} the value $\alpha_{SJ}=0.5$ was used. However, for such value
of $\alpha_{SJ}$ different values of $\varepsilon$ were needed for the correct 
description of the experimental data at moderate and high energies.
This problem was solved in \cite{SJ1}, where it was shown with the help
of more recent experimental data that all the data can be described with the
parameter values
\begin{equation}
\alpha_{SJ}\, =\, 0.9\;\; {\rm and} \quad \varepsilon\, =\, 0.024\,.
\end{equation} 

It is necessary to note that the process shown in Fig.~4c can be
realized very naturally in the quark combinatoric approach \cite{AnSh}
with the specified probabilities of a valence quark recombination
(fusion) with sea quarks and antiquarks.

\section{Comparison with the experimental data}

The probabilities $w_n$ in Eq.~(4) are calculated in the frame of Reggeon 
theory \cite{KTM}. The normalization constants $a_{\pi}$ (pion production), 
$a_K$ (kaon production), $a_{\bar{N}}$ ($B\bar{B}$ pair production), and $a_N$
(baryon production due to SJ diffusion) were determined  \cite{KTM,KaPi,Sh}
from the experimental data at fixed target energies, where the fragmentation 
functions are not constants. The values of these parameters have not been
modified for the present calculations, while the values of correspondent 
constants for hyperons have been calculated by quark combinatorics 
\cite{AnSh,CS}. For sea quarks 
we have
\begin{equation}
p:n:\Lambda + \Sigma:\Xi^0:\Xi^-:\Omega = 4L^3:4L^3:12L^2S:3LS^2:3LS^2:S^3
\,.
\end{equation} 
The ratio $S/L$ determines the strange suppression factor, and $2L+S = 1$.
Usually in soft processes the ratio $\lambda = S/L$ is assumed to be 0.2--0.35.
Inside this region it should be considered as a free parameter and in the 
numerical calculation we have used the value $\lambda = S/L = 0.25$ that leads 
to the best agreement with the data \cite{abe}. 

The calculated inclusive densities of different secondaries at RHIC, 
$\sqrt{s} = 200$ GeV, and LHC, $\sqrt{s} = 14$ TeV, energies are presented in 
Table 1, where one can see that the agreement of the QGSM calculations with 
RHIC experimental data \cite{abe} is reasonably good. 

\begin{center}
{\bf Table 1}
\end{center}

\vspace{5pt}

The QGSM results for midrapidity yields $dn/dy$ ($\vert y \vert < 0.5$) for 
different secondaries at RHIC and LHC energies. The results for
$\varepsilon = 0.024$ are presented only when different from the case 
$\varepsilon = 0$. \\

\begin{center}
\vskip 5pt
\begin{tabular}{|c||r|r|r|r|r|} \hline
Particle & \multicolumn{3}{c|}{RHIC ($\sqrt{s} = 200$ GeV)} & 
\multicolumn{2}{c|}{LHC ($\sqrt{s} = 14$ TeV)} \\ \cline{2-6}
& $\varepsilon = 0$ & $\varepsilon = 0.024$ & Experiment \cite{abe} & 
$\varepsilon = 0$ & $\varepsilon = 0.024$ \\   \hline
$\pi^+$         & 1.27    &        &                     & 2.54  & \\ 
$\pi^-$         & 1.25    &        &                     & 2.54  & \\
$K^+$           & 0.13    &        & $0.14 \pm 0.01$     & 0.25  &  \\
$K^-$           & 0.12    &        & $0.14 \pm 0.01$     & 0.25  &  \\
$p$             & 0.0755  & 0.0861 &                     & 0.177 & 0.184  \\
$\overline{p}$  & 0.0707  &        &                     & 0.177 &        \\
$\Lambda$       & 0.0328  & 0.0381 & $0.0385 \pm 0.0035$ & 0.087 & 0.0906 \\
$\overline{\Lambda}$ & 0.0304  &   & $0.0351 \pm 0.0032$ & 0.0867 &  \\
$\Xi^-$      & 0.00306  & 0.00359 & $0.0026 \pm 0.0009$ & 0.0108 & 0.0112 \\
$\overline{\Xi^+}$ & 0.00298 &     & $0.0029 \pm 0.001$  & 0.0108&  \\
$\Omega^-$       & 0.00020  & 0.00025 & * & 0.000902 & 0.000934 \\
$\overline{\Omega^+}$ & 0.00020  &    & * & 0.000902 &  \\
\hline
\end{tabular}
\end{center}

$^* dn/dy(\Omega^- + \overline{\Omega^+}) = 0.00034 \pm 0.00019$

\vspace{15pt}

The agreement of the QGSM calculations with RHIC experimental data \cite{abe}
is reasonably good.

The ratios of $\bar{p}/p$ production in $pp$ interactions at $\sqrt{s}=200$ 
GeV as the functions of rapidity have been calculated in the QGSM with 
the same parameters used in \cite{SJ2,SJ3}, and they are in reasonable 
agreement with the experimental data \cite{BRA} if the SJ contribution with  
$\varepsilon = 0.024$ is included, while the disagreement is evident for the 
calculation without SJ contribution (i.e. with  $\varepsilon = 0$). It is 
necessary to note that at asymptotically high  energies the ratio $\bar{p}/p$ 
in the central region is expected to be equal to the unity, so any deviation 
of the $\bar{p}/p$ ratio from unity has to be explained by some special 
reason. One can see in Table~1 that at the RHIC energies the SJ contribution 
makes the deviation of $\bar{p}p$ from unity in the midrapidity region about 
three times bigger than in the calculation without SJ contribution.

The QGSM predicts the deviation of $\bar{p}p$ ratios from unity due to SJ 
contribution on the level of 3-4\% accuracy even at the LHC energy. Without SJ 
contribution these ratios are exactly equal to unity.

The QGSM calculations \cite{SJ1} predict practically equal values of 
$\bar{B}/B$ ratios in midrapidity region independently on baryon strangeness, 
what is qualitatively confirmed by the RHIC data on Au-Au collisions 
\cite{MuNa}. In the case of $\Omega/\bar{\Omega}$ production in $pp$ 
collisions we obtain a non-zero asymmetry (i.e. more $\Omega$ than 
$\bar{\Omega}$), that is necessary absent in the naive quark model or in all 
recombination models, since both $\Omega$ and $\bar{\Omega}$ have no common 
valence quarks with the incident particles.

\begin{figure}[h]
\centering
\vskip -0.8cm
\includegraphics[width=.75\hsize]{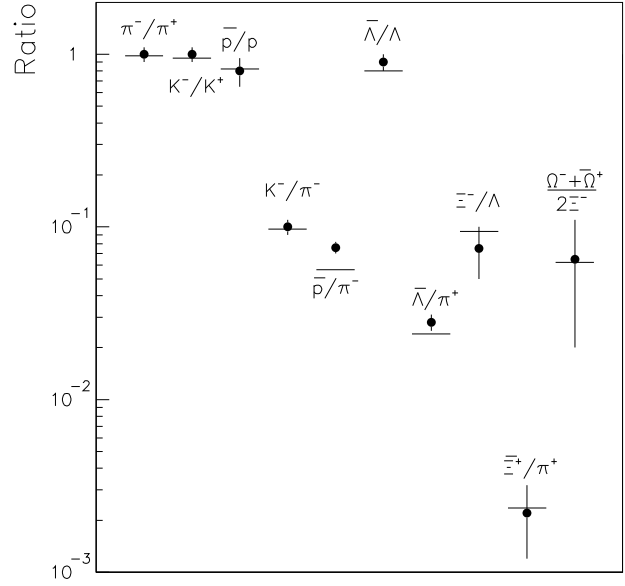}
\vskip -0.8cm
\caption{\footnotesize
Ratios of different secondaries produced in midrapidity region in $pp$ 
collisions at $\sqrt{s}=200\,$GeV. Short horizontal solid lines show results 
of the QGSM calculations.}
\vskip -0.3cm
\end{figure}

In Fig.~5 we reproduce the experimental data on ratios of yields of different 
secondaries \cite{abe} together with our calculations. Agreement is good 
except for only the point of the $\bar{p}/\pi^-$ ratio. From the comparison 
of our results with experimental data presented in Table~1 and Fig.~5 we can 
conclude that the universal parameter $\lambda$ = 0.25 describes the ratios of
$\Lambda/p$, $\Xi/\Lambda$, and $\Omega/\Xi$ production in a reasonable way.

\section{Conclusion}

We discuss the role of string junction diffusion in the baryon charge transfer 
over large rapidity distances for the cases of $pp$ collisions at RHIC and LHC 
energies. The inclusion of the SJ contribution provides a reasonable 
description of the main bulk of the existing experimental data. The 
calculations of the baryon/antibaryon yields and asymmetries without SJ 
contribution \cite{ACKS,SJ1} clearly diverge for most of the experimental
data, where this contribution should be important.
   Similar results for antibaryon to baryon production ratios at RHIC and LHC 
energies are presented in \cite{Bopp1}. They are obtained in the framework of 
DPMJET-III Monte Carlo. Some numerical difference with our results comes 
mainly from the different values of $\alpha_{sj}$ parameter.

{\bf Acknowledgements}

We are grateful to F. Bopp for useful discussions. 
This paper was supported by  Ministerio Educaci\'on y Ciencia of Spain under 
project FPA 2005--01963 and by Xunta de Galicia and, in part, by grants 
RFBR-07-02-00023 and RSGSS-1124.2003.2.

\end{document}